\def\cm{{\rm\thinspace cm}}

\def\erg{{\rm\thinspace erg}}

\def\keV{{\rm\thinspace keV}}
\def\ks{{\rm\thinspace ks}}
\def\km{{\rm\thinspace km}}

\def\Msun{\hbox{$\rm\thinspace M_{\odot}$}}

\def\s{{\rm\thinspace s}}
\def\yr{{\rm\thinspace yr}}

\def\ergps{\hbox{$\erg\s^{-1}\,$}}

\def\kmps{\hbox{$\km\s^{-1}\,$}}

\def\Msunpyr{\hbox{$\Msun\yr^{-1}\,$}}

\def\pcmsq{\hbox{$\cm^{-2}\,$}}

\def\psqcm{\hbox{$\cm^{-2}\,$}}

\documentstyle[psfig,times]{mn}
\begin{document}
\title{On the detectability of distant Compton-thick 
obscured quasars}
\author[]
{\parbox[]{6.in} {A.C.~Fabian$^1$, R.J.~Wilman$^2$ and C.S.~Crawford$^1$\\
\footnotesize
1. Institute of Astronomy, Madingley Road, Cambridge CB3 0HA \\
2. Sterrewacht Leiden, Postbus 9513, 2300 RA Leiden, The Netherlands }}

\maketitle
\begin{abstract}
Chandra and XMM--Newton have resolved the 2--8~keV X-ray Background
(XRB) into point sources. Many of the fainter sources are obscured
active galactic nuclei (AGN) with column densities in the range of
$10^{22}-10^{23}\pcmsq,$ some of which have quasar-like luminosities.
According to obscuration models, the XRB above 8~keV is dominated by
emission from Compton-thick AGN, with column densities exceeding
$1.5\times 10^{24}\pcmsq$. Here, we consider whether Compton-thick
quasars are detectable by Chandra and XMM--Newton by their direct
(i.e. not scattered) X-ray emission. Detectability is optimized if the
objects individually have a high luminosity and high redshift, so that
the direct emission has a significant flux in the observed band. Using
a simple galaxy formation model incorporating accreting black holes,
in which quasars build most of their mass in a Compton-thick manner
before expelling the obscuring matter, we predict that moderately deep
100~ks Chandra and XMM--Newton exposures may contain a handful of
detectable Compton-thick quasars. Deep Ms or more Chandra images
should contain 50--100 distant, optically-faint, Compton-thick
sources. In passing we show that radiation pressure can be as
effective in expelling the obscuring gas as quasars winds, and yields
a black hole mass proportional to the velocity dispersion of the host
bulge to the fourth power.
\end{abstract}


\section{Introduction}

Most of the 0.5--2~keV soft X-ray Background (XRB) has been resolved
by ROSAT (Hasinger et al 1999; Lehmann et al 2001) and recently most
of the harder 2--7~keV XRB has been resolved by Chandra (Mushotzky et
al 2000; Giacconi et al 2001; Barger et al 2001; Brandt et al 2001)
and XMM (Hasinger et al 2001). The source identifications are
incomplete at the present time but indicate that about one third of
the sources are associated with normal quasars, another third with
optically-bright galaxies and the last third with optically-faint
galaxies (Mushotzky et al 2000; Barger et al 2000; Alexander et al
2001). X-ray hardness ratios indicate that the fainter sources are
generally harder and consistent with absorbed sources. This confirms
obscuration models for the XRB (Setti \& Woltjer 1989; Madau,
Ghisellini \& Fabian 1994; Comastri et al 1995; Gilli, Salvati \&
Hasinger 2001).

Detailed studies, including X-ray spectroscopy of the brighter sources
(e.g. Crawford et al 2001) have however shown that the column
densities of absorbing material lie in the range of $10^{22}$ to a few
times $10^{23}\psqcm$. Such columns are insufficient to produce the
$\nu I_{\nu}$ peak in the XRB at about 30~keV (Marshall et al 1982).
In order to match this peak, sources with columns exceeding
$1.5\times 10^{24}\psqcm$ are required, i.e. sources which are Compton thick.

Here we address this issue and ask whether such sources should have
been observed in Chandra and XMM images, and whether their apparent
absence indicates a problem for the obscuration model of the XRB.
Compton-thick sources are certainly found in the local Universe, with
two of the nearest active galactic nuclei (AGN), the Circinus galaxy
and NGC4945 (Matt et al 2000, and references therein) being in this
class. Good examples of powerful Compton-thick AGN are NGC6240
(Vignati et al 1999) which is at low redshift and IRAS\,09104+4109
which is at redshift $z=0.442$ (Franceschini et al 2000; Iwasawa et al
2001).  The main issue is whether Compton-thick quasars are common.

Sources which are mildly Compton-thick ($10^{24}<N_{\rm
H}<10^{25}\psqcm$) show direct emission peaking above 10~keV. When
they are heavily Compton thick ($N_{\rm H}>10^{25}\psqcm$, this
emission peak is down-scattered by Compton recoil and is therefore
highly degraded. How much is seen in the 2--8~keV band from low
redshift Compton-thick objects depends almost completely on how much
emission is scattered into our line of sight. For example, the direct
line of sight to the well-known obscured AGN NGC\,1068 has a column
density exceeding $10^{25}\psqcm$ (Matt et al 1997), the X-ray
emission seen is all scattered flux. The $z=0.92$ hyperluminous IRAS
galaxy F15307, known to host a quasar from observed scattered
(polarised) optical broad lines, shows no X-ray flux detectable by
ASCA and ROSAT (Ogasaka et al 1997; Fabian et al 1996). The difference
between these sources is probably determined by the geometry of the
obscuring medium, which may be torus-like in the case of NGC\,1068 and
more spherical for F15307. Since synthesis models for the XRB spectrum
indicate that most (85 per cent) of the AGN power is absorbed (Fabian
\& Iwasawa 1999), a spherical geometry is more likely to be relevant
to the sources dominating the hard XRB than a torus one. The scattered
flux fraction is likely to be small. It is the absorbed direct flux
which we are interested in here.

Redshift does help to shift so render mildly Compton-thick quasars
detectable, leading to an inverse K-correction effect (Wilman \&
Fabian 1999). However the sources still need to be luminous to be
detectable. For the spectrum of the XRB, it is the mildly
Compton-thick objects which dominate at the $\nu I_{\nu}$ XRB peak.
The lowest redshift members of the Compton-thick class contribute most
to the highest energy part of that peak.

To assess whether Compton-thick quasars ought to be detected, we have
taken one simple obscured AGN model for the XRB (from Wilman, Fabian
\& Nulsen 2000; WFN), and predicted the number of Compton-thick
sources expected per square degree. We have then passed the predicted
spectra through the response curves of Chandra and XMM to predict the
number expected to be detected. Our results indicate that some
Compton-thick sources should be detectable, particularly in the
Chandra 1~Ms images (Alexander et al 2001; Rosati et al 2001). Chandra
does not have the collecting area above 8~keV to detect such hard
sources readily; XMM has too much internal background (relevant
because of its much larger PSF) at those energies. There may yet be
renewed claims to resolve the XRB when distant Compton-thick sources
are finally found in reasonable numbers. We argue that our results are
not strongly dependent on the model adopted.

\section{The WFN model}
For the sake of completeness, we provide here an outline of the WFN
model. At its heart is the semi-analytic galaxy formation code
developed by Nulsen \& Fabian~(1995), in which the Cole \&
Kaiser~(1988) block model is used to simulate the hierachical growth
of clustering, and the scheme of Nulsen \& Fabian~(1995) is used to
treat the behaviour of the gas within a collapsed halo. In short, gas
within a radius $R=R_{\rm{CF}}$, where the free-fall time is less than
the cooling time, rapidly forms stars, producing supernovae which can
then expel some (for normal galaxies) or all (in the case of dwarf
galaxies) of the remaining gas from the system. Any gas at
$R>R_{\rm{CF}}$ participates in a cooling flow (CF). Newly formed
normal galaxies are ellipticals unless all the hot gas is able to cool
before the present or the next hierarchical collapse, in which case
they are spirals. Any collapse with at most one infalling normal
galaxy forms a normal galaxy, with the stars of dwarf galaxies
contributing to the spheroid of the new system. Collapses with more
than one normal galaxy form a group or a cluster, as mergers between
normal galaxies are ignored.

Each of the smallest block model units contains a seed black hole of
mass $1.6 \times 10^{6}$\Msun, and when a block collapses all the
black holes associated with its merging subblocks are assumed to merge
into a single black hole at the centre of the new galaxy. Such nuclear
black holes are then fed by Bondi accretion of hot gas from the
cooling flow atmospheres which surround them, in accordance with the
model of Nulsen \& Fabian~(2000). Whilst thus accreting, the black
hole is assumed to radiate with an efficiency of 10 per cent, with a
2-10\keV~luminosity equal to 3 per cent of the bolometric one, with a
power law plus reflection spectrum. This intrinsic spectrum is then
absorbed by the isothermal distribution of cold dusty clouds deposited
by the cooling flow, as suggested by Fabian~(1999). The accretion is
terminated by wind-driven gas expulsion when the mass of the black
hole reaches a critical fraction of that in the surrounding spheroid,
thereby accounting for the observed correlation between the mass of
the remnant black hole and its host spheroid. Thereafter the object
shines as an optical quasar for $9 \times 10^{7}$\yr. For full details
of the model, see WFN.

\subsection{Wind or radiation pressure?}

The critical mass of the black hole was determined by invoking a wind
from the central engine (Fabian 1999; see also Silk \& Rees 1998 who
use an energy argument rather than the force one used here). We note
here that the radiation pressure of the absorbed radiation alone
yields a similar limit (i.e. if the force due to the radiation $\sim
L_{\rm abs}/c$, where the absorbed power $L_{\rm abs}\propto L_{\rm
Edd}$).

To pursue this in detail, we adopt the scenario of Fabian (1999),
where a fraction $f$ of the mass in the core of an isothermal galaxy
$M(<r)=2v^2r/G$ is in cold, X-ray absorbing gas. The gas within radius
$r_1$ has been accreted into a black hole of mass $M_{\rm BH}=f 2 v^2
r_1/G,$ leaving a column density $N=f v^2/2\pi G m_{\rm p} r_1$
beyond. Rearranging these formulae means that $$M_{\rm BH}=f^2v^4/G^2
\pi m_{\rm p} N.$$

Much of the radiation from the black hole is absorbed by the gas
beyond $r_1$ giving rise to a force on the column of $f_1 L_{\rm
abs}/4\pi r^2 c,$ where $f_1$ accounts for how much the matter traps
and reradiates the energy. $f_1\sim 0.5$ for an isolated cloud but may
be several for the envisaged optically-thick cloud surrounding the
nucleus. Then at the limit where the outward force due to radiation
balances the inward gravitational force on the column, $$L_{\rm
abs}=4\pi GMNm_{\rm p}cf_1^{-1}.$$ (The formula assumes that the
column is all at one radius; if it extends to $r_{\rm max}$ then a
factor of $\log(r_{\rm max}/r)$ enters into the inward force
expression.)

Let $L_{\rm abs}=f_2 L_{\rm Edd}=f_2 4\pi GMm_{\rm p}c/\sigma_{\rm T},$
where $\sigma_{\rm T}$ is the Thomson cross section.
Then $$N=f_1 f_2/\sigma_{\rm T}.$$
Substituting this into the above black hole mass formula then gives
$$M_{\rm BH}={{v^4\sigma_{\rm T}}\over{\pi G^2 m_{\rm
p}}}{f^2\over{f_1f_2}},$$
or $$M_{\rm BH}=2.3\times 10^9
{f^2\over{f_1f_2}}\left({v\over{200\kmps}}\right)^4.$$ This is the
mass of the black hole when the gas (which fuels it) is blown away.
The normalization changes slightly, by losing a factor of $f/\pi,$ if
$r_1$ is redefined as the Bondi accretion radius (as in WFN) for gas at
the galaxy virial temperature.

It compares well with the results of Ferrarese \& Merritt (2000) and
Gebhardt et al (2000) for reasonable values of $f,$ $f_1$ and $f_2$:
$$M_{\rm BH}=1.4\times 10^8
\left({f\over 0.25}\right)^2 \left({f_1 f_2}\right)^{-1}\left(
{v\over{200\kmps}}\right)^4\Msun.$$

The above result demonstrates that a wind is not essential for
expelling the gas surrounding a growing black hole. 

In the present model it appears more as if the galaxy potential
determines the black hole mass, rather than the other way round. Note
however that the gravitational binding energy of a galactic bulge,
where the velocity dispersion of the bulge is $300 v_{300}\kmps,$ is
$E_{\rm bulge}\approx 2\times 10^{-6}v_{300}^2 M_{\rm bulge} c^2$. The
energy from the central black hole $E_{\rm AGN}\approx 5\times
10^{-4}M_{\rm bulge} c^2$. So only one per cent of that energy can
have a major effect on the formation of that bulge and that effect may
have occurred when both the black hole and galaxy were young.

\section{The yield of Compton-thick sources}

For the above model with the free parameters set to the values given
in section 3 of WFN, we present in Figs.~1--4 results on the yield of
Compton-thick quasars in deep exposures with Chandra and XMM. Firstly,
Fig.~1 shows as a function of redshift the total areal density of
Compton-thick AGN in the model with intrinsic 2--10\keV~luminosities
in excess of $10^{44}$\ergps. As discussed by WFN, however, such
sources are obscured by $N_{\rm H} \sim 15 N_{\rm{T}}$ when the
obscured phase begins (where $N_{\rm T}=1.5 \times 10^{24}$\psqcm, the
column density above which a source becomes Compton-thick), and by
$N_{\rm H} \sim 3 N_{\rm{T}}$ at its end. Since there is no scattered
flux included in the model spectra and the direct emission is almost
completely suppressed for $N_{\rm H} > 10N_{\rm T}$ (especially
below 10\keV), the vast majority of the sources contributing to Fig.~1
will not be visible to Chandra or XMM. This is borne out by Figs.~2, 3
and 4, described below, which were produced by convolving the object
spectra in the observed frame with `response functions' giving the
effective area of the telescope and detector combination under
consideration.

In Fig.~2 we show the areal density as a function of redshift of both
the Compton-thick and Compton-thin AGN (the latter also includes the
unobscured quasars), which would produce more than 30 counts in the
5--10\keV~band in a 220\ks~exposure with XMM on the EPIC pn chip. This
is repeated in Fig.~3 for a 100~ks exposure. Fig.~4 shows the
analogous plot for the 2--8\keV~band with the Chandra ACIS S3 chip,
with the count threshold reduced to 10 counts owing to the smaller PSF
(and thus lower internal background) compared to XMM. The majority of
the detected objects contributing to the yields are quite close to the
count thresholds. For example, if the XMM threshold is increased to 40
counts, the yield of Compton-thick (Compton-thin plus unobscured) AGN
falls over all redshift by a factor 2--3 (1.5--2). Similarly,
increasing the Chandra detection threshold to 15 counts, results in a
fall by a factor of 2.5 in the yield of Compton-thick objects, whilst
the number of Compton-thin and unobscured objects are relatively
unchanged.

Most of the predicted Compton-thick sources should become detectable
in exposures of 1Ms (Fig. 5), such as are now being done with Chandra
(e.g. Alexander et al 2001; Rosati et al 2001). In an $8.4\times 8.4$
arcmin$^2$ region and 1~Ms exposure we predict 182 sources (Compton
thick and thin) in the 2--8~keV band whereas Alexander et al (2001)
find 102. We do not consider that this discrepancy in numbers is
serious, since most of the sources would be close to the detection
threshold. That many of the sources are seen in both soft and hard
energy bands (Alexander et al 2001) is also not a major problem since
a few per cent of the primary soft radiation may be scattered into our
line of sight.

The important issue of the redshift distribution of the sources is
considered in the next Section. where we also discuss the implications
of these results for the interpretation of deep Chandra and XMM
surveys.

\begin{figure}
\centerline{\psfig{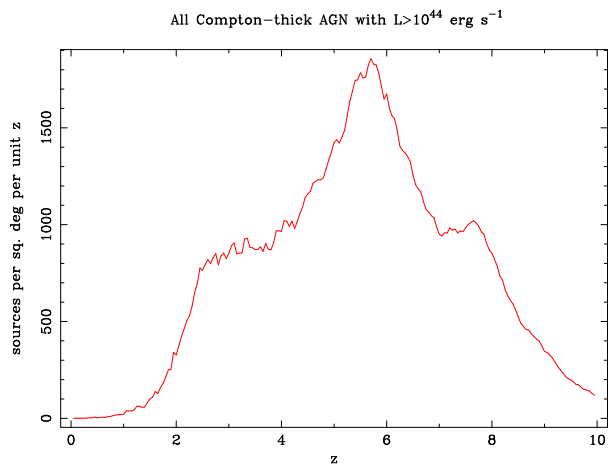}}
\caption{The areal density as a function of redshift of all Compton thick AGN in the model with intrinsic 2--10\keV~luminosity in excess of $10^{44}$\ergps, i.e. those of quasar luminosity.}
\end{figure}

\begin{figure}
\centerline{\psfig{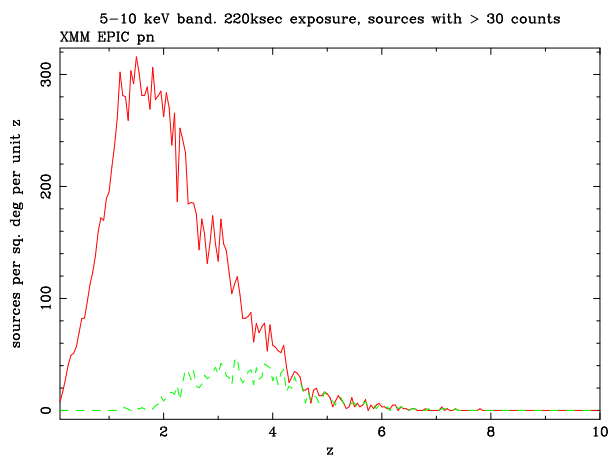}}
\caption{The solid line shows the areal density as a function of redshift of the Compton-thin and unobscured AGN combined (with no restriction on luminosity), showing more than 30 counts in the 5--10\keV band in the XMM EPIC pn chip in a 220\ks~exposure. The dashed line represents the yield of Compton-thick objects.}
\end{figure}

\begin{figure}
\centerline{\psfig{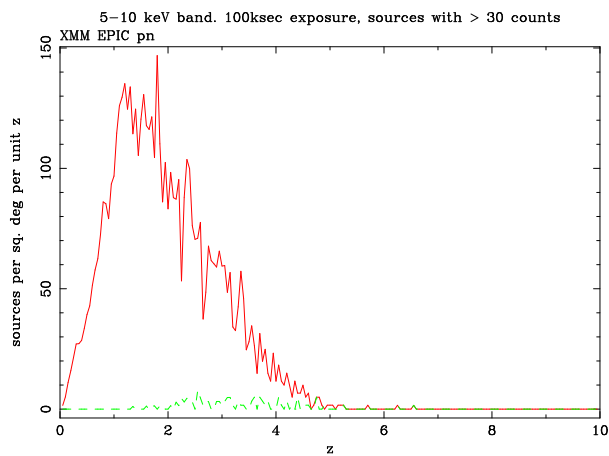}}
\caption{As in Fig.~2 but for an exposure time of 100\ks.}
\end{figure}

\begin{figure}
\centerline{\psfig{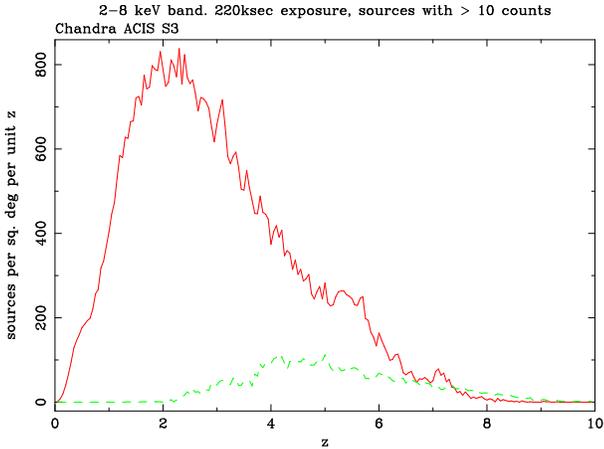}}
\caption{As in Fig.~2 but for the 2--8\keV~band on Chandra ACIS S3 chip, and a lower count threshold of 10 counts.}
\end{figure}

\begin{figure}
\centerline{\psfig{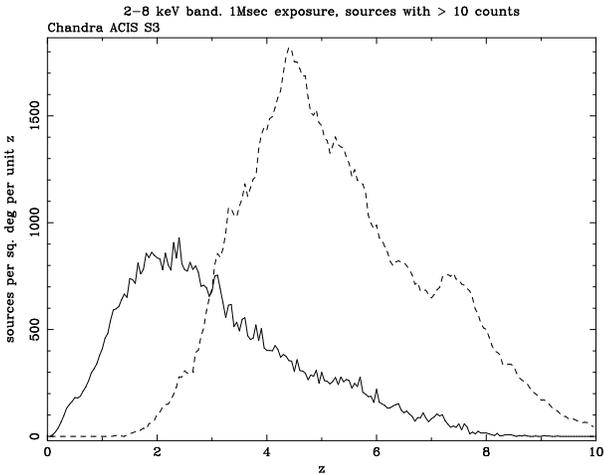}}
\caption{As in Fig.~2 but for the 2--8\keV~band, a 1~Ms exposure with
the Chandra ACIS S3 chip, and a lower count threshold of 10 counts.
The dashed line again shows the yield of Compton-thick objects.}
\end{figure}

\section{Discussion}
We see that the surface density of Compton-thick sources expected in a
220~ks exposure (equal in length to the initial Chandra observation of
the HDF-N and its vicinity by Hornschemeier et al.~2001) with Chandra
ACIS chip is about 200. A chip is $8\times8$ arcmin or 0.0178~sq~deg
which means that 3--4 faint hard sources are expected in such an
exposure. If the ACIS-I array is use, with 4 times the area,
then the number rises to about 10 (given
the degradation of sensitivity off axis: most of the photons detected
are above 6 keV). The sources have redshifts of 3--7 and are basically
detectable because of the inverse K-correction. Most of the
Compton-thick above $z=3$ become detectable in deeper exposures of
1~Ms.

In an XMM pn chip exposure of similar duration, concentrating on the
5--10~keV band, we have about 40 sources detected per square degree,
mostly in the redshift range of 2--5. The field of view is about 0.2
sq deg (30 arcmin diameter) leading to about 8 sources detected per
220~ks exposure, similar to the yield from Chandra. Source confusion
should not be a problem at such high photon energies.

The adopted WFN model thus predicts that the direct absorbed emission
from some Compton-thick quasars is detectable by Chandra and XMM with
similar yields. The absorbed primary emission from these sources will
be very hard and most will be detected close to the threshold for
detection (10 counts for Chandra, 30 for XMM), as discussed in section
2.1. Scattered emission may however allow the sources to be detected
also in softer energy bands. The appearance of such objects in the
optical and near infrared bands depends on the rate of star formation.
If this rate is low then they should appear as redshifted early-type
bulges, without strong emission lines, and if high they will be
brighter and of much later spectral type, probably with emission
lines.

So far only one source with a spectrum consistent with being Compton
thick has been reported (Norman et al 2001); its redshift $z=3.8$. The
observed X-ray emission in this case is not the absorbed direct
radiation but a reflected component. Only a guess can be made as to
the power of the direct component. It is this direct absorbed
component of such sources which is important for the spectral peak of
the XRB.

Our model predicts that more than half the sources detected in a Ms
exposure with Chandra will have redshifts ranging from 2--8. Most of
them would have $I$ band magnitudes exceeding 24, i.e. are optically
faint according to the definition of Alexander et al (2001). The
precise magnitudes would depend upon the star formation rate in the
host galaxy. If we assume that the obscured black hole growth phase
coincides with the major star formation phase in the galaxy, then the
star formation rate must be tens $\Msunpyr$. Then the $R$ (and
approximately $I$) magnitude of the hosts will be about 24 for objects
at $z\sim 3$ (e.g. Pettini et al 2001). Objects at $z\sim 6$ will be
about 1.5 mag fainter. Alexander et al (2001) have 47 objects fainter
than $I=24$ (15 with $I>25.3$), which is about half the number we
predict.

At this stage we consider that our model remains viable, with a factor
of two in number densities arrangeable by a small alteration of some
of the parameters (e.g. the gas fraction and/or the gas metallicity:
note that most of the sources are predicted to be close to the
detection threshold, even in 1Ms). Further work on the optically-faint
population in the Ms Chandra fields will resolve this issue by
comparison with the redshift distribution in Fig.~5.

We now ask how robust our conclusions might be, since they depend on a
specific model. First we note that they should be a reasonable (factor
of two) approximation to any model which has the obscured phase before
the 'normal', unobscured, quasar phase, since the model has that
emission peaking between $z=1-2$ as observed. It also produces a mean
local black hole density in agreement with that inferred from
observation (Merritt \& Ferrarese 2001). The higher redshift ($z>3$)
Compton-thick sources contribute 10--20 per cent of the intensity of
the XRB spectrum in the 5-12~keV band, and so are not of great
importance to the spectrum; the lower redshift sources with $1<z<2$
contribute far more and dominate the spectrum in the 20--30~keV band.
These last objects are completely undetectable with Chandra or XMM,
unless they have scattered soft emission. The lowest redshift
Compton-thick sources contribute most at the highest energies.

We conclude that distant Compton-thick quasars, which may dominate the
XRB intensity over the 5--20~keV band and represent the growth phase
of massive black holes, are just detectable with present instruments.
A handful may be found in 100~ks exposures at flux levels close to the
detection threshold. 50--100 distant Compton-thick sources are
predicted to be detectable in Chandra deep fields of 1~Ms or more. The
sources should be optically faint and probably appear as high redshift
starburst galaxies.

\section{Acknowledgements}
ACF and CSC thank the Royal Society for support.


\begin{thebibliography}{}
\bibitem []{}  Alexander D.M., Brandt W.N., Hornschemeier A.E.,
Garmire G.P., Schneider D.P., Bauer F.E., Griffiths R.E., 2001, AJ,
122,2156
\bibitem []{}  Barger A., Cowie L.L., Mushotzky R.F., Richards E.A., 2000,
AJ, 121, 662 
\bibitem []{}  Brandt W.N., et al 2001, AJ, 122, 1
\bibitem []{}  Cole S., Kaiser N., 1988, MNRAS,233, 637  
\bibitem []{}  Comastri A., Setti G., Zamorani G., Hasinger G., 1995,
A\&A, 296, 1
\bibitem []{}  Crawford C.S., Fabian A.C., Gandhi P., Wilman R.J.,
Johnstone R.M., 2001, MNRAS, 324, 427
\bibitem []{}  Fabian A.C., Cutri R.M., Snith H.E., Crawford C.S.,
Brandt W.N., 1996, MNRAS< 283, L95 
\bibitem []{}  Fabian A.C., Iwasawa K., 1999, MNRAS, 303 L4
\bibitem []{}  Fabian A.C., 1999, MNRAS, 308, L39
\bibitem []{}  Ferrarese L., Merritt D., 2000, ApJ, 539, L9
\bibitem []{}  Gebhardt K. et al, 2000, ApJ, 539, L13
\bibitem []{}  Giacconi R., et al, 2001, ApJ, 551, 624
\bibitem []{}  Gilli R., Salvati M., Hasinger G., 2001, A\&A, 366, 407
\bibitem []{}  Hasinger G., Burg R., Giacconi R., Schmidt M., Trumper J.,
Zamorani G., 1998, A\&A, 329, 482
\bibitem []{}  Hasinger G., et al, 2001, A\&A, 365, L45
\bibitem []{}  Hornschemeier A., et al 2001, ApJ, 554, 742
\bibitem []{}  Lehmann I., 2001, ApJ, 371, 833
\bibitem []{}  Madau P., Ghisellini G., Fabian A.C., 1994, MNRAS, 270, L17
\bibitem []{}  Marshall F.E., et al 1980, ApJ, 253, 377
\bibitem []{}  Matt G., et al 1997, A\&A, 325, L13
\bibitem []{}  Matt G., Fabian A.C., Guainazzi M.,  Iwasawa K., Bassani L.,
Malaguti G., 2000, MNRAS, 318, 173
\bibitem []{}  Merritt D., Ferrarese L., 2001, MNRAS, 320, L30
\bibitem []{}  Mushotzky R.F., Cowie L.L., Barger A.J., Arnuad K.A., 2000,
Nature, 404, 459
\bibitem []{}  Norman C., et al 2001, ApJ submitted, (astro-ph/0103198)
\bibitem []{}  Nulsen P.E.J., Fabian A.C., 1995, MNRAS, 277, 561
\bibitem []{}  Ogasaka Y. et al 1997, PASJ, 49, 179
\bibitem []{} Pettini M. et al  2001, ApJ, 554, 981
\bibitem []{} Rosati P. et al 2001, ApJ in press (astro-ph/0110452)
\bibitem []{}  Setti G., Woltjer L., 1989, A\&A, 224, L21
\bibitem []{}  Silk J., Rees M.J., 1998, A\&A, 331, L1
\bibitem []{}  Vignati P., et al 1999, A\&A, 349, L57
\bibitem []{}  Wilman R.J., Fabian A.C., 2000, MNRAS, 309, 862 
\bibitem []{}  Wilman R.J., Fabian A.C., Nulsen P.E.J., 2000, MNRAS, 319,
583

\end{thebibliography}
\end{document}